\documentclass[11pt]{article}
\usepackage{moriond,epsfig}

\def\Journal#1#2#3#4{{#1} {\bf #2}, #3 (#4)}

\def\NIMA{{\em Nucl. Instrum. Methods} A}
\def\NPB{{\em Nucl. Phys.} B}
\def\PLB{{\em Phys. Lett.}  B}
\def\PRL{\em Phys. Rev. Lett.}

\def\be{\begin{equation}}
\def\ee{\end{equation}}
\def\bea{\begin{eqnarray}}
\def\eea{\end{eqnarray}}

\begin{document}
\vspace*{4cm}
\title{Recent results from K2K}

\author{ Taiki IWASHITA \\ {\it for the K2K collaboration}}

\address{KEK(High Energy Accelerator Research Organization),\\
1-1 Oho, Tsukuba, Ibaraki, Japan}

\maketitle\abstracts{
 Results from K2K-I and status of K2K-II (run after the SK accident in
November 2001) was reported.
 The results from K2K-I are the $\nu_{\mu}$ disappearance,
$\nu_e$ appearance, and studies of neutrino interaction with water nuclei.
 For K2K-II, steady increase of POT (number of Protons On Target) and number SK events
are shown. 
Also, a new front detector SciBar was installed from K2K-II.
}

\section{Introduction of K2K}\label{sec:intro}

K2K (KEK to Kamioka) experiment is the first long base-line neutrino
oscillation experiment using accelerator in the world~\cite{bib:k2k}.
The motivations of the experiment are: 

\begin{itemize}
\item to confirm $\nu_{\mu}$ disappearance
\item to search $\nu_{e}$ appearance
\item to study $\nu$ interaction at sub GeV region.
\end{itemize}

The 12 GeV PS provides about $6 \times 10^{12}$ protons per spill, in every 2.2 seconds.
Proton beam is extracted and bent in the direction of the far detector Super-Kamiokande, 
and is injected on the target-horn system.
K2K has two magnetic horns to focus the produced $\pi^+$.
The focused charged pions decay into $\nu_{\mu}$ in the 200m of decay section.

\begin{figure}
\begin{center}
\psfig{figure=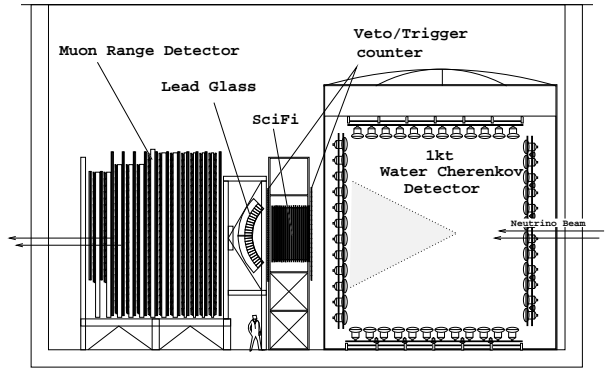,height=3in}
\caption{ A drawing of K2K-I near detectors.
\label{fig:nd}}
\end{center}
\end{figure}

In the 300m downstream of the target, we have the near detector complex system
to measure neutrino flux and energy spectrum right after their generation.
The near detector system consists of a 1kton water Cherenkov detector (1KT),
a scintillating fiber detector (SciFi)~\cite{bib:scifi}, a lead glass calorimeter (LG),
a muon range detector (MRD)~\cite{bib:mrd} in K2K-I period (Jun.1999 $\sim$ July 2001).
In K2K-II (Jan.2003 $\sim$ ), LG was removed and a new scintillator bar detector
(SciBar)~\cite{bib:scibar} was installed alternatively.
SciFi, LG, SciBar, MRD are called fine grain detector (FGD) collectively.
A drawing of K2K-I near detectors is shown in Figure~\ref{fig:nd}.

The far detector of K2K experiment is the Super-Kamiokande (SK)
water Cherenkov detector which has 50kton of fiducial mass.
After an accident in Nov.2001, we had rebuilt the detector successfully
and resumed data taking in Dec.2002.
This is called SK-II.
It has about 5200 20 '' PMTs with acrylic covers in the inner detector,
1885 8 '' PMTs in the outer detector.

\section{$\nu_{\mu}$ disappearance}\label{sec:numu}

\begin{figure}
\begin{center}
\psfig{figure=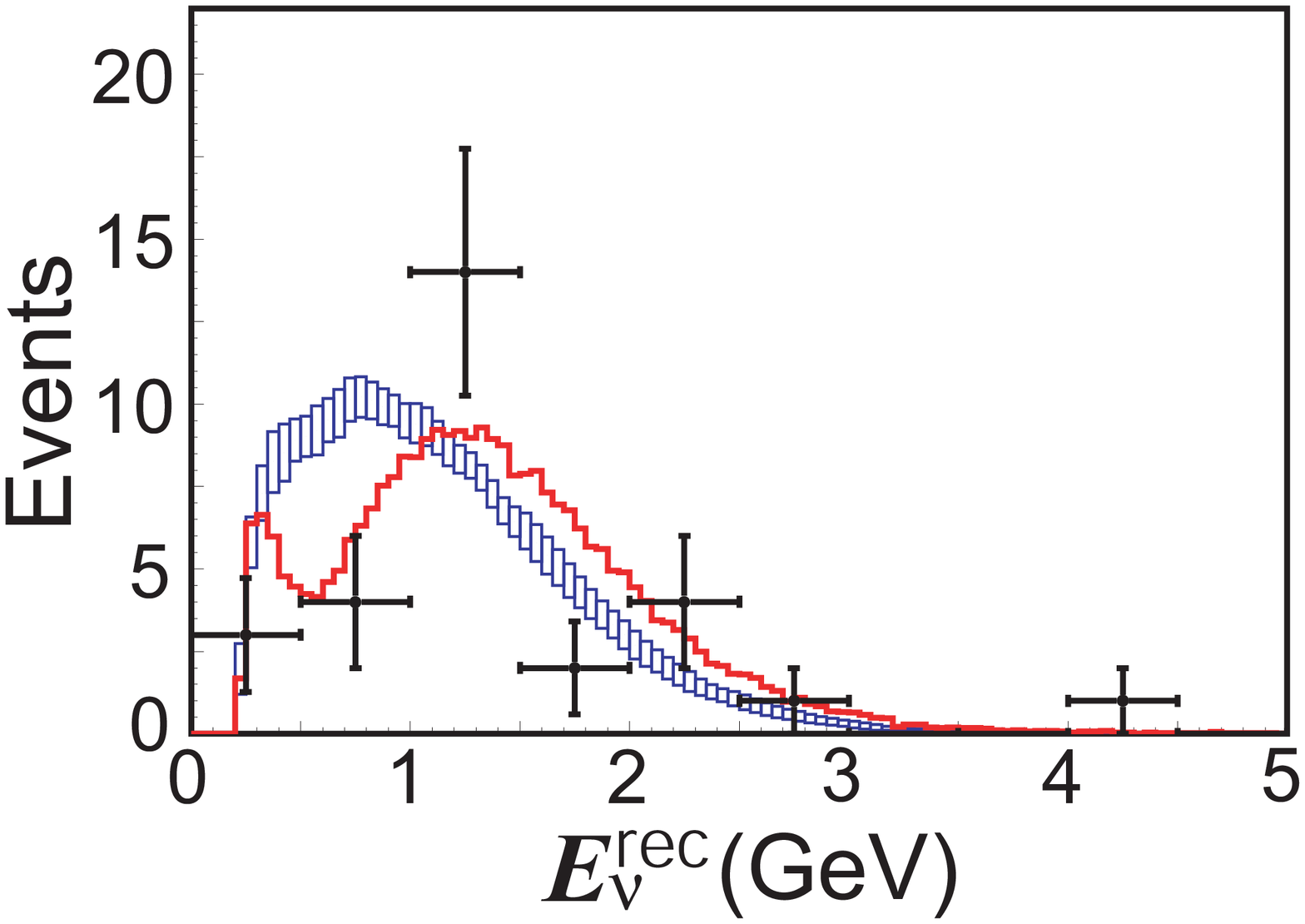,height=2.in}
\psfig{figure=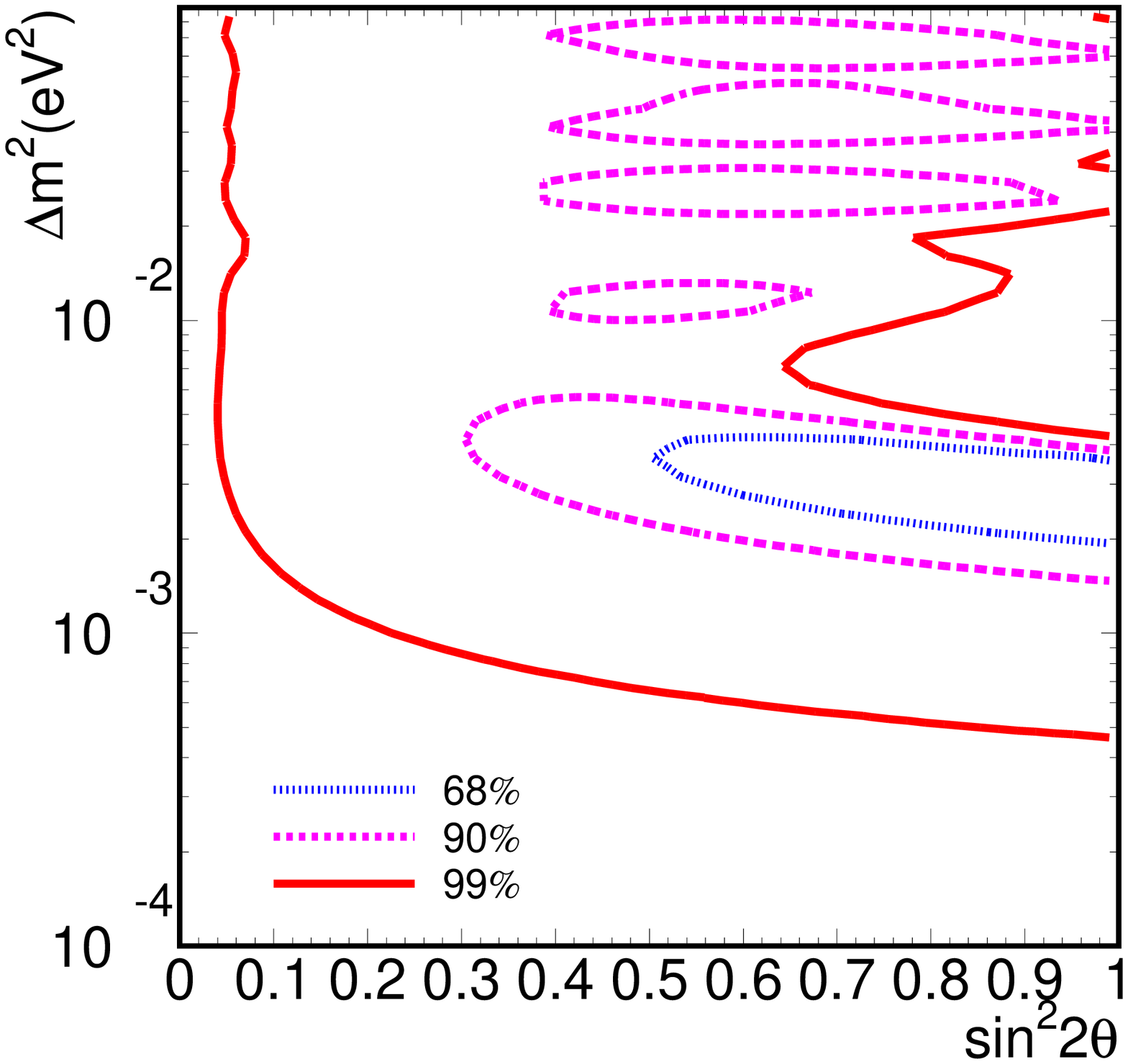,height=2.5in}
\caption{
(left) The reconstructed neutrino energy spectrum at SK (dots)
and expected ones from the near detector (box histogram: without oscillation.
open histogram: with best fit oscillation parameters).
(right) The allowed region of two oscillation parameters (sin$^2 2 \theta$, $\Delta m^2$) 
\label{fig:cont}}
\end{center}
\end{figure}

In K2K-I, we observed 56 events at SK while the expectation from the near
detector is $80.1^{+6.2}_{-5.4}$. Figure \ref{fig:cont} (left) is the reconstructed 
neutrino energy spectrum at SK (dots) and expected ones from the near detector
(box histogram: without oscillation. open histogram: with best fit oscillation parameters).
The null oscillation probability is less than 1\% from an analysis using both the 
number of events and the energy spectrum shape.
The allowed region of two oscillation parameters (sin$^2 2 \theta$, $\Delta m^2$) 
is shown in Figure \ref{fig:cont} (right).

\begin{figure}
\begin{center}
\psfig{figure=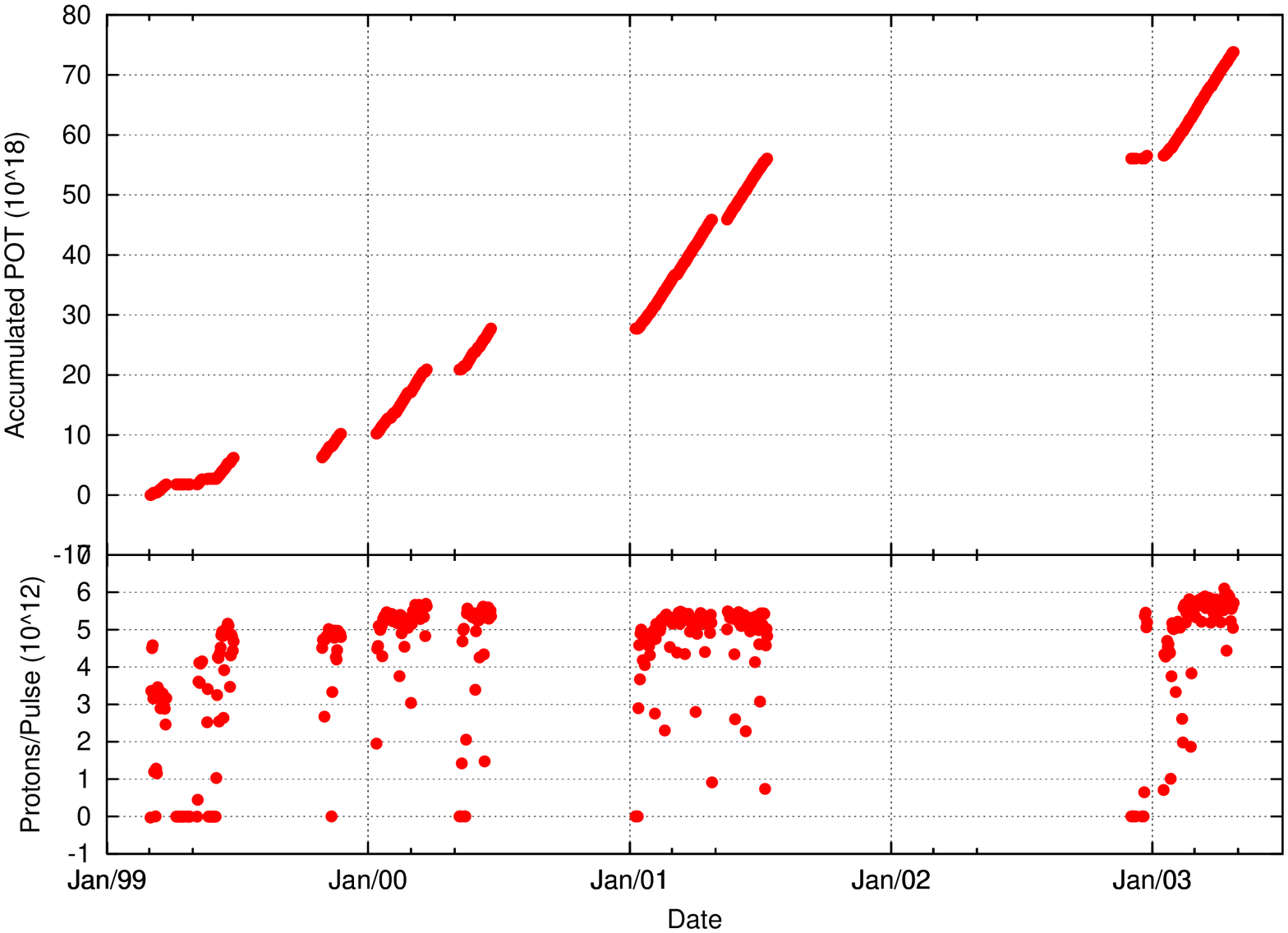,height=2.5in}
\psfig{figure=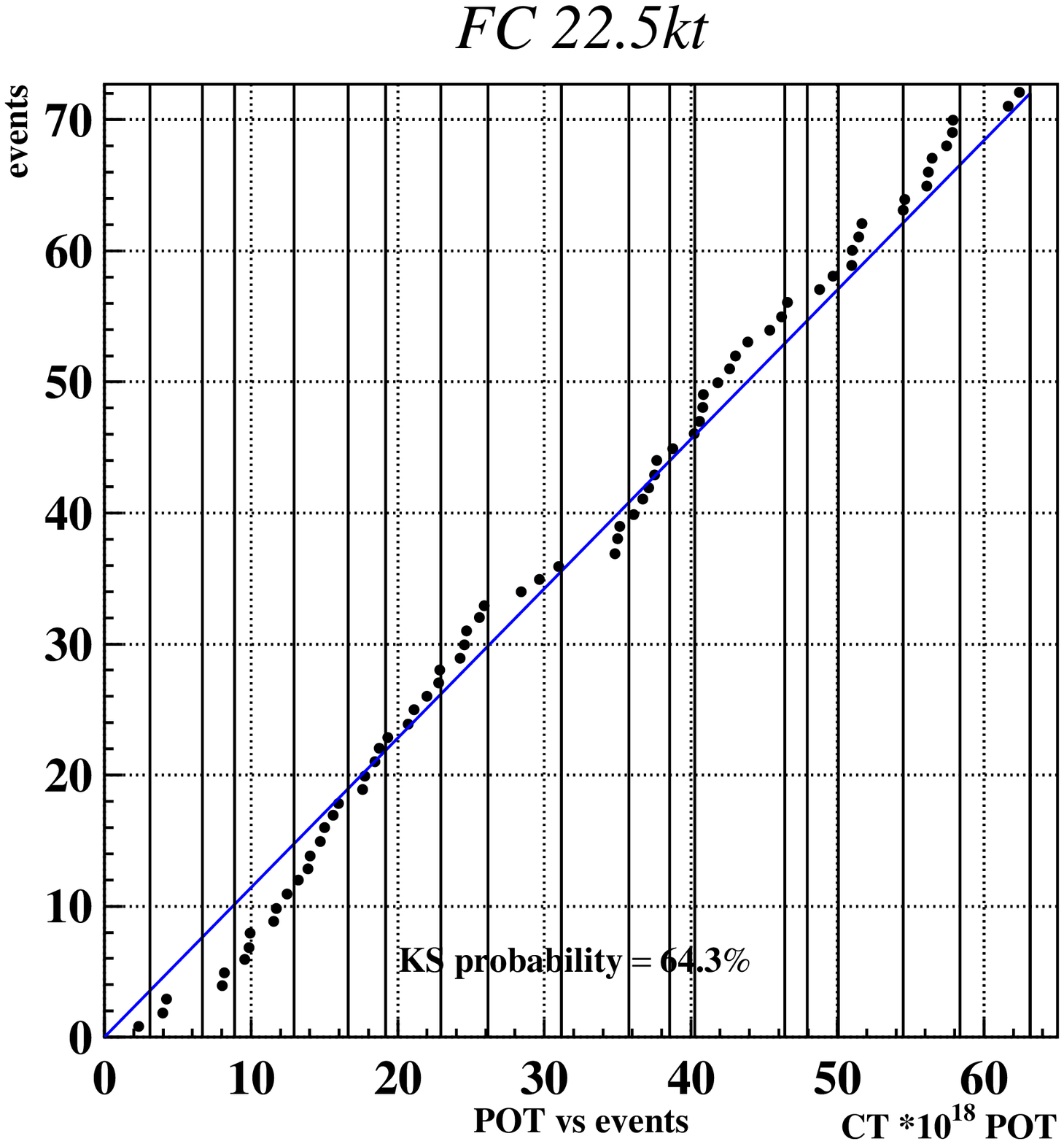,height=2.5in}
\caption{
(left) Delivered protons on target (POT) in K2K-I (Jun.1999 - Jul.2001)
and K2K-II (Jan.2003 - ).
(right)
The increasing SK events in K2K-I and K2K-II.
\label{fig:pot}}
\end{center}
\end{figure}

Figure \ref{fig:pot} (left) shows delivered protons on target (POT).
After an accident of SK, we re-started neutrino beam run from 
Jan. 2003 and  POT is successfully increasing.
The increasing SK events are plotted in Figure \ref{fig:pot} (right).
Sixteen new SK events are found in K2K-II by May 2003.
Data accumulation in SK is stable and the event rate is consistent
with K2K-I.

\subsection{New near detector SciBar}\label{subsec:scibar}

SciBar detector consists of $\sim$ 15000 channels of extruded scintillator bar
with WLS (wave length shifter) fiber readout.
It's fully working from Oct. 2003, collecting about
7000 neutrino event candidates per month.

The detector has high efficiency for short track,
and p/$\pi$ separation by using the light yield is expected.
Therefore the SciBar has high efficiency for 2-track CCQE
(charged current quasi-elastic) events with low backgrounds.

\section{$\nu_e$ appearance}\label{sec:nue}

We searched CCQE ( $\nu_{e}$ +n $\rightarrow$ e + p ) interaction
at SK in K2K-I data.
To select them, single ring is required because the proton momentum is typically below
the Cherenkov threshold at K2K beam energy.
For the next cut, we defined two PID; 
one is by ring pattern, a ring image from an electron is diffused
by the electro magnetic shower.
Another one use the opening angle of the Cherenkov ring.
An Opening angle of Cherenkov ring from a low energy muon is small because of mass.

\begin{figure}
\begin{center}
\psfig{figure=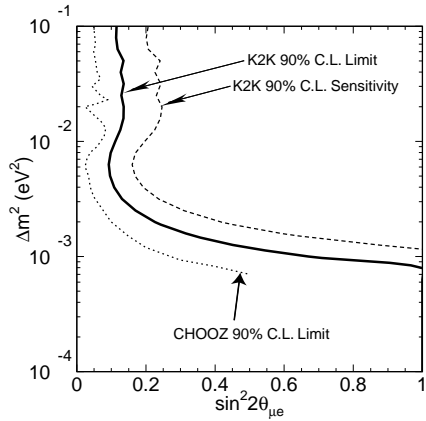,height=2.5in}
\caption{ The exclude region of $\nu_e$ appearance in the parameter space.
The dotted line is the CHOOZ excluded region, converted into
sin$^2 2\theta$, assuming full mixing in the 2-3 sector.
\label{fig:nue_cont}}
\end{center}
\end{figure}

Table \ref{tab:nue} gives a reduction summary of the $\nu_e$ disappearance search.
Starting from 56 fully contained SK events, 
only 1 $\nu_{e}$ event candidate was found after the selection.
The main background comes from the $\nu_{\mu}$ interactions, especially
$\pi^0$ production is dominant.
The observed 1 event is consistent with the expected background.
The exclude region of $\nu_e$ appearance in the parameter space is shown
in Figure \ref{fig:nue_cont}. 
The dotted line on the figure is the CHOOZ ($\bar{\nu_e}$ disappearance)
excluded region, converted into
sin$^2 2\theta_{\mu e}$, assuming full mixing in the 2-3 sector.

\begin{table}[t]
\caption{Reduction summary of the $\nu_e$ appearance search in K2K-I.
In the background estimation by $\nu_{\mu}$ MC, null oscillation is assumed.
``Signal $\nu_{e}$ MC(CC)'' means the expected number of events in
sin$^2 2\theta$=1, $\Delta m^2 = 2.8 \times 10^{-3} eV^2$.
\label{tab:nue}}
\vspace{0.4cm}
\begin{center}
\begin{tabular}{|c c c c c|}
\hline
                &Data & $\nu_{\mu}$ MC & Beam $\nu_{e}$ MC & Signal $\nu_{e}$ MC(CC)\\
\hline\hline
FCFV            & 56  & 80 & 0.82 & 28 \\
Single ring     & 32  & 50 & 0.48 & 20 \\
PID(e-like)     &  1  & 2.9& 0.42 & 18 \\
$E_{vis}$ 100MeV&  1  & 2.6& 0.41 & 18 \\
w/o decay-e     &  1  & 2.0& 0.35 & 16 \\
\hline
\end{tabular}
\end{center}
\end{table}

This result is submitted to Physical Review Letters. \cite{bib:nue}

\section{Neutrino interaction studies}\label{sec:ma}

\subsection{$\pi^0$ cross section measurement at 1KT}

In our $\nu_e$ analysis, neutral current $\pi^0$ production
(NC1$\pi^0$) is a main background. 
NC1$\pi^0$ is also used to check the possibility of
$\nu_{\mu} \rightarrow \nu_s$.
It will be more important in near future T2K experiment.

Because a neutral pion decays to 2 gamma, 
NC1$\pi^0$ candidates are selected by requiring fully contained 2 ring e-like.
With K2K-I 1KT data, the cross section ratio of NC1$\pi^0$ and CC (charged current)
interaction was measured~\cite{bib:pi0};
$ \sigma( \rm{NC}1\pi^0) / \sigma( \rm{\nu_{\mu}CC}) = 0.063 \pm 0.001 (stat.) \pm 0.006 (sys.). $

\subsection{$M_A$ measurement at FGD}

Preliminary analysis of the axial-vector mass ($M_A$)~\cite{bib:ma}
of quasi-elastic interaction.
We used data for $Q^2 \ge$  0.2 (GeV/$c^2$) only to avoid large uncertainty
of the region due to the large nuclear effects.
The coherent pion and multi-pion interactions have also uncertainty.
As a result of $\chi^2$ fitting, the measured $M_A$ value of QE
with water target is $1.06 \pm 0.03 (stat.) \pm 0.14$ (preliminary).

\section*{Summary}
 From K2K-I, the probability of null oscillation case is less than 1\%
with the analysis using the number of events and the spectrum shape both.
For $\nu_e$ appearance analysis, one $\nu_e$ candidate was found in the data set
but it's consistent with background. The exclude region for $\nu_e$ appearance is set.

After the accident at SK, K2K-II data taking started from Oct.2003.
The event rate is consistent with K2K-I run. The new detector SciBar was fully
installed and is working from the beginning of K2K-II.

With K2K-I data, some neutrino interaction studies are going on.
1KT measured the cross section ratio of NC1$\pi^0$ and CC (charged current)
interaction was measured;
$ \sigma( \rm{NC}1\pi^0) / \sigma( \rm{CC}) = 0.063 \pm 0.001 (stat.) \pm 0.006 (sys.). $
FGD measured the $M_A$ of quasi-elastic interaction; the measured $M_A$
with water target is $1.06 \pm 0.03 (stat.) \pm 0.14$ (preliminary).

\end{document}